\documentclass[10pt,conference]{IEEEtran}
\IEEEoverridecommandlockouts
\usepackage[nobreak,space,compress]{cite}
\usepackage{comment}
\usepackage{etoolbox}  

\usepackage{booktabs}
\usepackage{enumitem}
\usepackage{graphicx}

\usepackage{amsmath,amssymb,amsfonts}
\usepackage{textcomp}
\usepackage[table]{xcolor}  
\usepackage{tcolorbox}
\def\BibTeX{{\rm B\kern-.05em{\sc i\kern-.025em b}\kern-.08em
    T\kern-.1667em\lower.7ex\hbox{E}\kern-.125emX}}
\usepackage{ulem}
\usepackage{listings}
\usepackage[hyphens]{url}
\usepackage{hyperref}
\usepackage{caption}
\usepackage{algorithm}
\usepackage{algpseudocode}
\usepackage{multirow}
\usepackage{subfig}
\usepackage{balance}

\newcommand{\ourtitle}{Self-Admitted Technical Debt in LLM Software: An Empirical Comparison with ML and Non-ML Software}

\begin{document}
\title{\ourtitle}

\author{
    \IEEEauthorblockN{Niruthiha Selvanayagam}
    \IEEEauthorblockA{École de Technologie Supérieure - ÉTS Montreal\\
    Montréal, Canada\\
    niruthiha.selvanayagam.1@ens.etsmtl.ca}
    \and
    \IEEEauthorblockN{Taher A. Ghaleb}
    \IEEEauthorblockA{Trent University\\
    Peterborough, Canada\\
    taherghaleb@trentu.ca}
    \and
    \IEEEauthorblockN{Manel Abdellatif}
    \IEEEauthorblockA{École de Technologie Supérieure - ÉTS Montreal\\
    Montreal, Canada\\
    manel.abdellatif@etsmtl.ca}
}

\maketitle

\begin{abstract}
Self-admitted technical debt (SATD), referring to comments in which developers explicitly acknowledge suboptimal code or incomplete functionality, has received extensive attention in machine learning (ML) and traditional (Non-ML) software. However, little is known about how SATD manifests and evolves in contemporary Large Language Model (LLM)-based systems, whose architectures, workflows, and dependencies differ fundamentally from both traditional and pre-LLM ML software. 
In this paper, we conduct the first empirical study of SATD in the LLM era, replicating and extending prior work on ML technical debt to modern LLM-based systems. We compare SATD prevalence across LLM, ML, and non-ML repositories across a total of 477 repositories (159 per category). We perform survival analysis of SATD introduction and removal to understand the dynamics of technical debt across different development paradigms.
Surprisingly, despite their architectural complexity, our results reveal that LLM repositories accumulate SATD at similar rates to ML systems (3.95\% vs. 4.10\%). However, we observe that LLM repositories remain debt-free 2.4x longer than ML repositories (a median of 492 days vs. 204 days), and then start to accumulate technical debt rapidly.
Moreover, our qualitative analysis of 377 SATD instances reveals three new forms of technical debt unique to LLM-based development that have not been reported in prior research: \textit{Model-Stack Workaround Debt}, \textit{Model Dependency Debt}, and \textit{Performance Optimization Debt}. Finally, by mapping SATD to the stages of LLM development pipeline, we observe that debt concentrates significantly higher in the deployment and pretraining stages. 
\end{abstract}

\section{Introduction}

Technical debt, the hidden cost of shortcuts in software development, has challenged software development since Cunningham first coined the metaphor in 1992~\cite{cunningham1992wycash}. When developers write comments such as ``\texttt{TODO: fix this properly}'' or ``\texttt{HACK: temporary workaround}'', they explicitly acknowledge these shortcuts. Such self-admitted technical debt (SATD) comments serve as signals~\cite{potdar2014satd,maldonado2015detecting}, marking where future maintenance is likely needed. Prior work has shown that SATD predicts maintenance challenges~\cite{li2022tdmanagement}, guides refactoring decisions~\cite{zazworka2013comparing}, and reveals developer concerns about code quality~\cite{xavier2020beyond}. 

As machine learning (ML) became widespread, researchers discovered new SATD patterns, including experimental pipelines and data dependencies, unlike traditional (non-ML) software, where technical debt primarily stems from structural and design issues~\cite{sculley2015hidden,bhatia2025empirical}.
Since ChatGPT's release in November 2022, Large Language Model (LLM) development has introduced fundamentally new architectural patterns that differ from both ML and traditional (non-ML) systems~\cite{openai_chatgpt}. Modern LLM applications embed logic in prompts, use vector stores in Retrieval-Augmented Generation (RAG), coordinate multi-model workflows via agent frameworks, and depend on volatile external APIs~\cite{lewis2020retrieval, survey2025modern, ragsurvey2024}.
These architectural shifts have introduced new forms of technical debt. For example, at the prompt layer, minor wording changes can break entire workflows~\cite{hiddenTechDebtLLM}, while orchestration frameworks may obscure critical business logic across distributed components~\cite{langchainProblems}. At the infrastructure layer, token-based pricing can introduce economic debt that might compound with scale~\cite{hiddenCostsRAG}, and dependency on rapidly changing external APIs can create model lock-in risks~\cite{techDebtAI}.
Though practitioners encounter these issues daily~\cite{zhang2024reliability}, systematic empirical evidence remains limited.

Most existing SATD research is based on data collected before the widespread adoption of LLMs, with datasets~\cite{potdar2014satd,xavier2020beyond,bhatia2025empirical,obrien2022shades} that do not capture modern LLM architectural patterns. For example, Bhatia \textit{et al.}~\cite{bhatia2025empirical} conducted an empirical analysis of SATD in ML and non-ML systems and provided a taxonomy of such debts without considering LLM-based systems. In addition, even recent work on LLM technical debt~\cite{aljohani2025promptdebt} has examined only single-model prompt invocation and does not capture modern multi-agent, multi-model, or RAG-based workflows, nor does it examine the evolution of SATD over time or compare LLMs against ML and non-ML software systems.
In this paper, we conduct the first empirical study of SATD in the LLM era. We replicate and extend Bhatia \textit{et al.}'s study~\cite{bhatia2025empirical} by constructing a new curated dataset of 159 LLM repositories (developed since November 2022) and comparing them against 159 ML and 159 non-ML repositories from Bhatia \textit{et al.}'s dataset, using the latest updates to their metadata, commit history, and code comments. We analyze over five million comment events to identify patterns in the introduction, persistence, and removal of SATD across various development paradigms.

Our experiments show that, despite their architectural complexity, LLM repositories accumulate SATD at similar rates to ML ones (3.95\% vs. 4.10\%) but remain debt-free $2.4$ times longer than ML repositories (a median of 492 days vs. 204 days) before debt starts to accumulate rapidly.
We identify three new SATD categories specific to LLMs, accounting for 22.4\% of the classified debt: \textit{Model-Stack Workaround Debt} (6.3\%), \textit{Model Dependency Debt} (11.2\%), and \textit{Performance Optimization Debt} (4.9\%).
When mapped to LLM development pipeline~\cite{hu2024llmcharacterization}, we observe that debt is most concentrated in \textit{Deployment and Monitoring} (30\%) and \textit{Pretraining} (23\%) stages.
Our analysis shows that foundational SATD (the first debt introduced in a file) is rarely removed (removal rates below 5\%), 
with developers tending to clean more recent, localized debt while leaving foundational issues intact.
The first SATD introduced into any file remains nearly permanent (only 4.2\% ever removed in LLM repositories) despite 49.1\% overall comment-level removal. Our model achieves 82\% accuracy for LLM debt versus 96\% for ML debt. This gap suggests that while ML debt remains highly predictable from traditional software metrics (confirming Bhatia \textit{et al.}~\cite{bhatia2025empirical}), LLM debt involves new factors (prompt engineering, RAG orchestration, API volatility) that conventional commit-level features cannot fully capture.

This paper makes the following contributions:

\begin{enumerate}[leftmargin=1.1cm, topsep=0pt, itemsep=2pt]    
    \item
We present the first empirical comparison of SATD in LLM, ML, and non-ML software, revealing that LLM development has unique debt accumulation patterns.
    
    \item
    We extend the existing SATD taxonomy~\cite{bhatia2025empirical} by identifying three new technical debt types specific to LLM software.
    
    \item
    We provide insights on where debt accumulates (deployment $>$ pretraining $>$ infrastructure), what predicts its persistence, and how debt removal differs across paradigms.
    
    \item
    We release our complete replication package~\cite{our_replication_package}, including analysis scripts, our extended SATD detector, and the curated dataset of 159 LLM repositories with full git histories for replication and future studies.
\end{enumerate}

The remainder of this paper is organized as follows.
Section~\ref{sec:Related_Work} reviews related work.
Section~\ref{sec:Research_Questions} presents our research questions.
Section~\ref{sec:Methodology} details our methodology.
Section~\ref{sec:Results} reports results across five research questions.
Section~\ref{sec:Threats} discusses threats to validity.
Section~\ref{sec:Implications} presents implications of our findings.
Section~\ref{sec:Conclusion} concludes the paper and suggests directions for future work.

\section{Related Work}
\label{sec:Related_Work}

This section reviews prior research on self-admitted technical debt, examining its evolution from traditional software systems through machine learning applications to the emerging patterns in LLM-based development.

\noindent\textbf{Self-Admitted Technical Debt in Traditional Systems.} SATD was first characterized by Potdar and Shihab~\cite{potdar2014satd}, who showed that developers explicitly acknowledge suboptimal or temporary design choices in comments. Subsequent work expanded SATD detection and classification in traditional software systems. For instance, Maldonado and Shihab~\cite{maldonado2015detecting} refined SATD categorization, while Xavier \textit{et al.}~\cite{xavier2020beyond} demonstrated that SATD extends beyond code, appearing in issue trackers and developer discussions. Studies on the consequences of SATD revealed that it is tightly coupled with maintenance effort and change-proneness~\cite{zazworka2013comparing, li2022tdmanagement}. Large-scale datasets such as those by Lenarduzzi \textit{et al.}~\cite{lenarduzzi2021sonarqube} and studies using static analysis tools~\cite{guaman2017sonarqube,singh2017pmd} further highlight the prevalence of such debts. 

\vspace{2pt}
\noindent\textbf{Technical Debt in Machine Learning Systems.} ML systems introduce forms of technical debt that go beyond those found in traditional systems. Sculley \textit{et al.} first highlighted how ML systems accumulate \emph{data dependency debt}, \emph{configuration debt}, and \emph{boundary erosion}, noting that seemingly small design shortcuts can have widespread consequences through data pipelines and model behavior~\cite{sculley2015hidden}. Building on this theoretical framework, Bhatia \textit{et al.}~\cite{bhatia2025empirical} conducted a landmark empirical study comparing SATD in 318 ML and 318 non-ML repositories, finding that ML systems contain twice the SATD density of traditional (non-ML) software. They found ML repositories introduce debt $140$ times faster but also remove it $3.7$ times faster. Their work established the empirical foundation for understanding ML technical debt at scale. O'Brien \textit{et al.}~\cite{obrien2022shades} further expanded this understanding by identifying \emph{23 distinct types} of ML-specific technical debt through qualitative analysis. Together, these studies underscore that ML systems carry structural, data-centric, and operational debt patterns that differ significantly from traditional systems, with Bhatia \textit{et al.}'s quantitative findings providing the empirical basis that motivates our replication and extension to LLM systems.

\vspace{2pt}
\noindent\textbf{Emerging Technical Debt in LLM-Based Development.}
Empirical research on LLM-specific SATD is still nascent. Aljohani and Do~\cite{aljohani2025promptdebt} examined SATD in early LLM API usage patterns and found that prompt design, rather than code structure, is a primary source of LLM-specific TD. However, their focus on single-model prompt invocation does not capture modern multi-agent, multi-model, or RAG-based workflows, nor does it examine the evolution of SATD over time.

\vspace{2pt}
\noindent\textbf{Need for Contemporary SATD Replication.}
Prior SATD research has been conducted mainly on systems predating the widespread adoption of LLMs and modern ML pipelines. As development practices shift toward data-centric, prompt-centric, and orchestration-heavy architectures, the assumptions underlying classical SATD studies may no longer hold. Existing work lacks an empirical comparison of SATD behaviors across LLM, ML, and traditional repositories. Understanding these differences is critical for both research and practice: it informs SATD detection methods, provides insight into the maintainability of AI-driven projects, and updates the broader TD literature for a new generation of software systems. This work contributes to filling this gap by providing the first SATD lifecycle analysis across these three categories of repositories.

\section{Research Questions}
\label{sec:Research_Questions}
This study replicates and extends Bhatia \textit{et al.}'s~\cite{bhatia2025empirical} SATD analysis of ML systems to explore how SATD manifests in LLM systems. We examine whether architectural shifts in LLM development, such as prompt engineering, external APIs, and multi-model orchestration, change technical debt patterns compared to traditional ML and non-ML software. Accordingly, we compare the prevalence, types, stages, and survival of SATD in LLM, ML, and non-ML systems. Our study is structured
 around five research questions that mirror and extend those of Bhatia \textit{et al.}~\cite{bhatia2025empirical}.

\vspace{3pt}
\noindent\textbf{RQ1. What is the prevalence of SATD in LLM-based systems compared to ML and non-ML systems?}
Bhatia \textit{et al.}\cite{bhatia2025empirical} found that ML repositories contain twice as much SATD as non-ML repositories, attributing this to experimental workflows and high code churn. LLM development introduces additional complexity through prompt engineering, external API dependencies, and orchestration frameworks. We investigate whether these characteristics increase the prevalence of SATD in such systems or whether matured AI practices have reduced the accumulation of technical debt.

\vspace{3pt}
\noindent\textbf{RQ2. What are the different types of SATD in LLM-based systems?}
The original study extended Bavota and Russo's SATD taxonomy~\cite{bavota_russo} to include ML-specific categories like configuration debt and inadequate testing. LLM development introduces new components absent in classical ML, such as RAG orchestration, vector databases, human-in-the-loop workflows, prompt templates, embedding models, token-cost optimization, context-window management, and agent orchestration. These elements may create new debt categories within complex multi-model pipelines. We explore whether LLM repositories exhibit distinct SATD patterns that require further taxonomy extensions beyond pre-LLM ML systems.

\vspace{3pt}
\noindent\textbf{RQ3. Which stages of the LLM pipeline are more prone to SATD?}
Bhatia \textit{et al.}~\cite{bhatia2025empirical} found that model building and data preprocessing stages accumulate the most debt in classical ML pipelines. We investigate whether this holds for LLM systems or if debt instead concentrates in new architectural components (prompt orchestration, agentic logic, and RAG/vector database integration) or in LLM-specific stages (like pretraining, fine-tuning, or deployment and monitoring).

\vspace{3pt}
\noindent\textbf{RQ4. How long does SATD survive in LLM-based systems?}
The original study found that ML projects introduce SATD 140 times faster but also remove it 3.7 times faster than non-ML projects. In LLM systems, the coupling between prompts, models, and external services may affect both SATD introduction and removal. We investigate whether the SATD lifecycle differs in LLM repositories due to API volatility and architectural constraints.

\vspace{3pt}
\noindent\textbf{RQ5. What are the characteristics of long-lasting SATDs in LLM-based systems?}
Bhatia \textit{et al.}~\cite{bhatia2025empirical} found that long-lasting SATD in ML systems arises from large code changes spanning multiple low-complexity files. LLM repositories may differ due to their reliance on configuration files, prompt templates, and external service integrations. We investigate whether the factors driving persistent SATD vary across LLM, ML, and non-ML repositories.

\section{Methodology}
\label{sec:Methodology}

\subsection{Replication Scope and Goals}

To study SATD in LLM systems, we replicate and extend Bhatia \textit{et al.}'s study of SATD in ML systems~\cite{bhatia2025empirical}. Their original analysis examined SATD in 318 ML and 318 non-ML Python projects using SATD detection and survival analysis to characterize SATD prevalence, types, and lifecycle.
Our goals are twofold:
\emph{Replication}: Reproduce their main quantitative findings on SATD prevalence and survival in ML and non-ML repositories using their replication package, detector, and analysis pipeline, adapted to current software versions, as the original replication package required updates to deprecated library dependencies;
\emph{Extension}: Extend the study to contemporary, post-ChatGPT LLM repositories by constructing a curated dataset of LLM projects and applying the same SATD detection and survival analysis, as well as a refined SATD taxonomy tailored to LLM-specific development practices.

We follow the original study’s design wherever possible, including project curation criteria, comment extraction, SATD detection, and statistical tests, and clearly document deviations, such as the time span analyzed (temporal window), LLM-specific search queries, and taxonomy extensions.

\subsection{LLM Dataset Curation}

Similar to Bhatia \textit{et al.}~\cite{bhatia2025empirical}, we collect balanced repository sets with complete project evolution history via GitHub API to retrieve active systems, following established MSR practices~\cite{kalliamvakou2016,munaiah2017}. We construct 41 search queries combining LLM-specific terms (``chatbot'', ``gpt'', ``openai'', ``anthropic''), frameworks (``langchain'', ``llama-index''), and architectural markers (``rag'', ``embedding'', ``agent'') with quality filters adapted from Bhatia \textit{et al.}: (C1) non-forked repositories, (C2) $\,\geq 5\,$ Python files, (C3) $>1$ month of development history, and (C4) presence of LLM frameworks. Data collection was completed in early November 2025.
A multi-signal classification pipeline confirmed LLM relevance through (1) README semantics, (2) metadata/topic labels, and (3) dependency inspection. We organize 159 LLM repositories into a two-layer taxonomy: \emph{Infrastructure/Tools} (n=101) and \emph{End-User Applications} (n=58), with functional tags including Evaluation/Testing, Agentic, Training/Fine-tuning, Serving/Inference, RAG, General Chatbot, and Prompt Engineering. Our dataset targets post-November 2022 repositories to capture contemporary LLM constructs (e.g., prompt engineering, RAG, agentic orchestration, API-driven invocation), distinct from Bhatia \textit{et al.}'s pre-LLM dataset (2015-2021), enabling faithful SATD replication within the LLM era.

\subsection{ML and Non-ML Repository Selection}

Though Bhatia \textit{et al.}~\cite{bhatia2025empirical} analyzed 318 ML and 318 non-ML Python projects, we sample 159 repositories per category to match our LLM dataset as in Bavota and Russo~\cite{bavota_russo}. Their work, which forms the basis for SATD classification in both Bhatia \textit{et al.}'s and our study, showed that 159 repositories provide sufficient statistical power for SATD analysis while remaining computationally feasible for manual classification. This sample size ensures balanced comparison across LLM, ML, and non-ML repositories, without overrepresenting any type.
For ML repositories, we randomly select 159 projects from Bhatia \textit{et al.}'s replication package\footnote{\url{https://github.com/Aaditya-Bhatia/Self-Admitted-Technical-Debt-in-Machine-Learning-Software}}, ensuring coverage across their six domains (Image Processing, Natural Language Processing (NLP), Audio Processing, Autonomous Gameplay, Self-Driving Cars, Other ML). For non-ML repositories, we replicate their GitHub API methodology using the keywords "python projects" and "server" with identical filtering criteria (non-forked, $\geq$5 Python files, $>$1 month history, no ML components), yielding 159 repositories that matched the ML dataset characteristics.

\subsection{SATD Detection Pipeline}

We employ Liu \textit{et al.}'s NLP-based SATD detector~\cite{liu2018satd}, a state-of-the-art SATD classification tool that achieved 0.82 F1-score in prior evaluations. The detection process involved three steps: (1) extracting Python comments with Comment-Parser, (2) applying the pre-trained classifier to identify SATD instances, and (3) tracking temporal metadata for survival analysis.
For RQ1, we analyze SATD prevalence in the current codebase (502,592 unique comments in LLM repositories). For RQ4-RQ5, we track 3,713,429 comment events (additions and removals) across all commits in LLM repositories, enabling both cross-sectional prevalence assessment and longitudinal lifecycle tracking.

\subsection{SATD Sampling and Classification Procedure}

To address RQ2, we select a representative subset of SATD comments from the dataset obtained in RQ1. Using stratified sampling across the two layers with a 95\% confidence level and 5\% margin of error, we obtain a random sample of 377 SATD instances. This ensures balanced domain representation and reduces bias toward any single application area. 
Before coding the sampled comments, two co-authors jointly examined an initial set of 70 SATD instances (distinct from the 377 used in the final analysis) to align on the classification scheme. Coding followed a consensus-based approach and a card-sorting process~\cite{snyder2003paper}: both authors discussed each instance, resolving uncertainties through collaborative negotiation.
We report raw agreement: 60 of the 70 instances (85.7\%) were coded without disagreement, while 10 instances (14.3\%) required further discussion before reaching consensus. Through this iterative discussion, the two co-authors also identified patterns that did not fit existing categories and collectively defined three new LLM-specific debt types. This collaborative approach ensured consistent interpretation while allowing for emergent category discovery, which is particularly important given the novel characteristics of LLM-related technical debt.

After that, we categorize the 377 sampled SATD comments using a card-sorting approach guided by the extended taxonomy proposed by Bavota and Russo~\cite{bavota_russo}\footnote{Bhatia \textit{et al.}~\cite{bhatia2025empirical} further adapted this taxonomy by adding \textit{Configuration Debt} and \textit{Inadequate Testing}, both of which we incorporate in our analysis.}. Card sorting facilitated iterative comparison, refinement of category boundaries, and consistent labeling based on the initial joint examination. During coding, we encountered SATD comments that contained only the keyword ``TODO'' with little or no explanatory description of the intended action. We labeled such instances as \emph{Undefined}, while those lacking context for technical debt were labeled as \emph{Unknown}. Out of the 377 sampled instances, 44\% (166) were identified as false positives. From the remaining 211 true SATD instances, we excluded those labeled as \emph{Undefined} or \emph{Unknown} , resulting in a final set of 143 relevant instances for RQ2. If an SATD instance did not fit the existing taxonomy, we introduce a new category, following Bhatia \textit{et al.}'s approach of extending Hindle \textit{et al.}'s taxonomy for ML systems~\cite{hindle2008what}.

\subsection{SATD Prevalence Metrics}

We employ two complementary normalization strategies to measure SATD prevalence, following established practices from Bavota and Russo~\cite{bavota_russo} and Bhatia \textit{et al.}~\cite{bhatia2025empirical}:

\vspace{3pt}
\noindent\textbf{Comment-Level Prevalence (Density).} This metric measures the proportion of developer comments that acknowledge technical compromise, indicating the concentration of technical debt within the project's documentation, as follows:

\begin{equation}
    \text{SATD}_{\text{density}} = \frac{\text{Total SATD comment instances}}{\text{Total comment instances}}
\end{equation}

\vspace{3pt}
\noindent\textbf{File-Level Prevalence (Diffusion).} This metric measures the spread of technical debt across the codebase, indicating how widely SATD affects the project's architecture as follows:

\begin{equation}
\text{SATD}_{\text{diffusion}} = \frac{\text{Files with } \geq 1 \text{ SATD}}{\text{Total Python files}}
\end{equation}

These metrics offer complementary perspectives: density reflects the intensity of debt accumulation, while diffusion shows its architectural reach.

\subsection{Statistical Analysis Methods}

We employ several statistical techniques to ensure robust comparisons across repository types.

\vspace{3pt}
\noindent\textbf{Wilcoxon Rank-Sum Test:} We use the Wilcoxon rank-sum test (Mann-Whitney U test)~\cite{wilcoxon1945} to compare SATD prevalence between repository types. This non-parametric test is suitable for our non-normally distributed SATD percentages with varying sample sizes and is robust to outliers~\cite{mann_whitney_1947}.

\vspace{3pt}
\noindent\textbf{Effect Size (Cohen's d):} While p-values indicate statistical significance, Cohen's $d$~\cite{cohen1988} quantifies the practical magnitude of differences. We interpret effect sizes following Cohen's conventions: $d < 0.2$ (negligible), $0.2 \leq d < 0.5$ (small), $0.5 \leq d < 0.8$ (medium), $d \geq 0.8$ (large). This distinction is crucial, as large sample sizes can yield statistical significance even for trivially small differences~\cite{sullivan2012}.

\vspace{3pt}
\noindent\textbf{Bonferroni Correction:} To control the family-wise error rate in multiple pairwise comparisons (LLM vs. ML, LLM vs. Non-ML, and ML vs. Non-ML), we apply Bonferroni correction~\cite{dunn1961multiple}, lowering the significance threshold $\alpha$ from 0.05 to 0.017 for three comparisons. We prioritize controlling false positives to ensure that reported differences between repository types reflect genuine patterns rather than statistical artifacts from multiple testing."

\vspace{3pt}
\noindent\textbf{Survival Analysis:} For temporal analyses (RQ4), we employ Kaplan-Meier estimation~\cite{kaplan_meier_1958} to model SATD introduction and removal times, with log-rank tests~\cite{mantel1966} for comparing survival curves across repository types. This approach properly handles censored data, which occurs when SATD remains unresolved at the study endpoint and we observe only a minimum survival time rather than the actual removal time.

\section{Results and Discussion}
\label{sec:Results}

\subsection{RQ1: SATD Prevalence Across Repository Types}

Table~\ref{tab:satd_prevalence} presents SATD prevalence results (density) measured as the ratio of SATD comments to total comments. We found 3.95\% SATD in LLM repositories, 4.10\% in ML repositories, and 3.23\% in non-ML repositories.

\begin{table}[h]
\centering
\caption{SATD prevalence in current codebase}
\vspace{-4pt}
\label{tab:satd_prevalence}
\begin{tabular}{lrrrr}
\hline
\textbf{Repository} & \textbf{Repositories} & \textbf{Total~~~} & \textbf{SATD~~~} & \textbf{SATD~~~} \\
\textbf{Type} & \textbf{Analyzed~~} & \textbf{Comments} & \textbf{Comments} & \textbf{Prevalence} \\
\hline
    LLM    & 159 & 502,592 & 19,875 & 3.95\% \\
    ML     & 159 & 153,602 &  6,293 & 4.10\% \\
    Non-ML & 159 & 120,791 &  3,896 & 3.23\% \\
    \hline
\end{tabular}
\vspace{-2pt}
\end{table}

The prevalence difference between ML and non-ML repositories is $1.27x$, compared to $2x$ as reported by Bhatia \textit{et al.}~\cite{bhatia2025empirical}. Wilcoxon rank-sum tests confirm significant differences between all repository pairs (all $p < 0.005$), exceeding the Bonferroni-corrected significance threshold of $\alpha = 0.017$ for three comparisons. Effect sizes indicate moderate practical significance for ML vs. non-ML (Cohen's d = 0.42) and LLM vs. non-ML (d = 0.38) comparisons, but minimal difference between ML-LLM repositories (d = 0.04). 
The comparable SATD prevalence between LLM (3.95\%) and ML (4.10\%) repositories challenges our initial hypothesis that LLM systems would accumulate substantially more technical debt due to their architectural complexity, external API dependencies, and fast-paced ecosystems. This similarity (d = 0.04) suggests that LLM development has benefited from lessons learned in earlier ML systems, adopting disciplined practices from inception rather than repeating the experimental chaos that characterized early ML projects. 

However, this similarity must be interpreted considering the different temporal contexts of our repository cohorts. The ML repositories, sampled from Bhatia \textit{et al.}'s 2015-2021 dataset, have had several years to mature and potentially undergo cycles of debt accumulation and removal by the time of our analysis. In contrast, LLM repositories are relatively young (post-2022), capturing them in their initial development phase. The post-2022 LLM ecosystem has been characterized by a high-pressure, fast-moving development, with teams racing to achieve state-of-the-art results and capture market share. This environment incentivizes shortcuts and quick iterations that would bias the LLM cohort toward accumulating more technical debt in shorter time frames.

\vspace{3pt}
\noindent\textbf{Practical Implication.}
The statistically similar levels of technical debt in LLM and ML projects (d = 0.04) reflect two opposing forces: the stabilizing effect of mature ML practices and the rapid “gold rush” pace of LLM development. The comparable SATD levels may indicate LLM projects accumulate debt faster, matching ML prevalence within 2–3 years, or that modern LLM development benefits from established ML practices. The narrowing gap between ML and non-ML repositories ($1.27x$ vs. $2x$ in 2021~\cite{bhatia2025empirical}) likely signals broader AI maturation, supported by standardized frameworks such as PyTorch Lightning and Hugging Face Transformers.
\vspace{-1pt}

\begin{figure}

    \centering
    \includegraphics[width=1\linewidth]{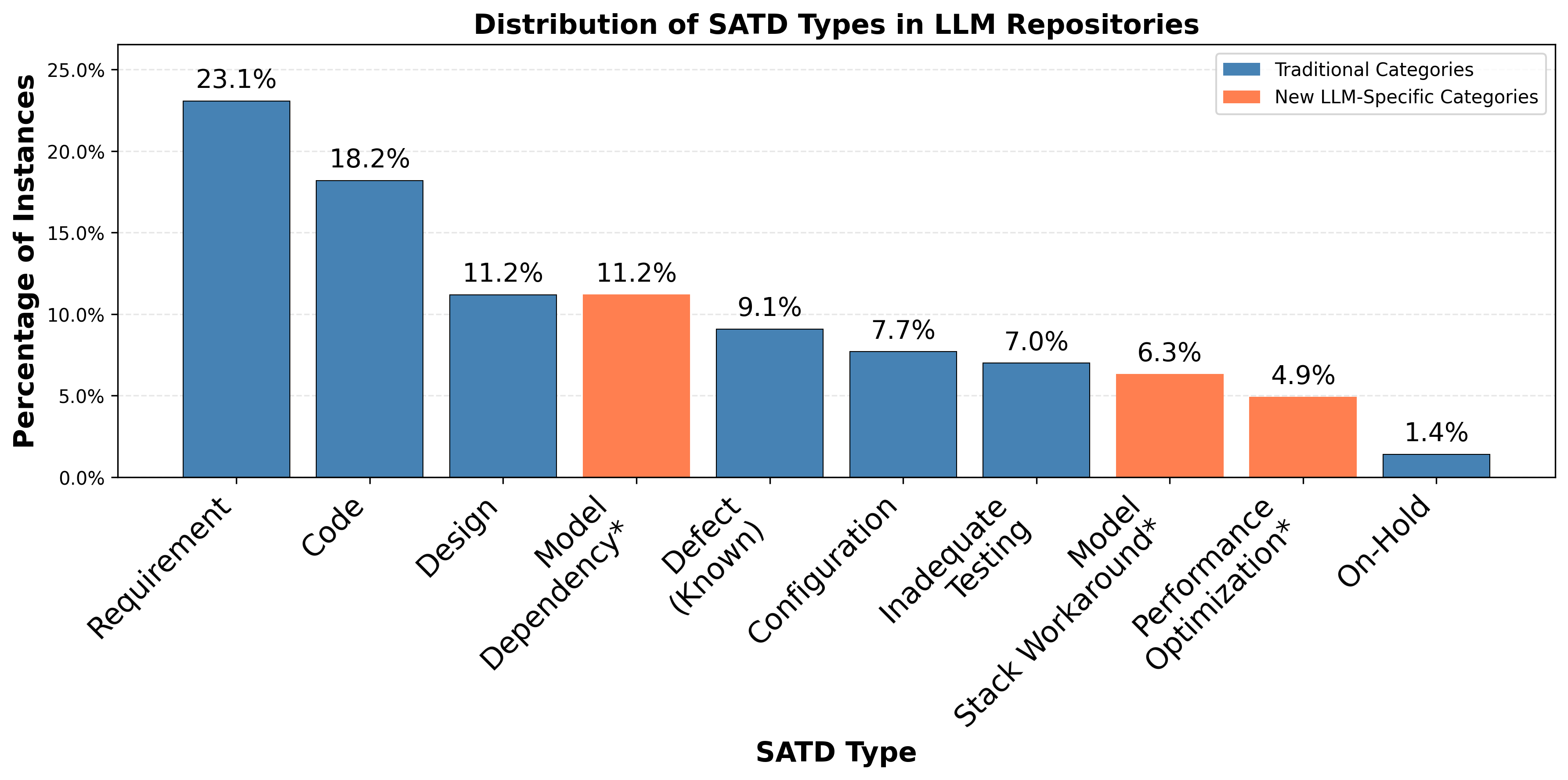}
    \caption{\textit{Distribution of SATD types in LLM repositories.}}
    \label{fig:satd_distribution}
    \vspace{-10pt}
\end{figure}

\begin{tcolorbox}[colback=gray!10, colframe=gray!50, boxrule=0.5pt, left=2pt, right=2pt, top=2pt, bottom=2pt]
\textbf{Answer to RQ1:} LLM and ML repositories show comparable SATD prevalence (3.95\% vs. 4.10\%) despite being much younger. This suggests that, given the temporal difference and development pressure on LLM projects, rapid debt accumulation in LLM systems is offset by benefits from mature ML tooling and practices.
\end{tcolorbox}

\subsection{RQ2: Types of SATD in LLM-based Systems}

We analyze 377 SATD instances from our LLM repositories to identify debt categories and extend existing taxonomies. Figure~\ref{fig:satd_distribution} presents the distribution of SATD types, revealing that although traditional categories remain dominant, 15\% of classified SATD comprises novel LLM-specific patterns.

\subsubsection{Extending Existing Taxonomies for Modern AI Systems}

Our analysis shows that existing SATD taxonomies do not adequately capture debt in LLM systems. Bhatia \textit{et al.}~\cite{bhatia2025empirical} extended Bavota and Russo's taxonomy~\cite{bavota_russo} with two ML-specific categories (Configuration Debt, Inadequate Testing), but these remain insufficient for LLM systems. The original taxonomy assumes object-oriented architectures (e.g., poor encapsulation), which fits Bhatia \textit{et al.}'s OO-heavy ML repositories but not modern LLM systems that rely on procedural logic, distributed runtimes, and hardware-specific pipelines~\cite{llm_architectures_survey_2024}. This architectural shift required substantial taxonomy expansion: although Bhatia \textit{et al.} made only minor ML-specific additions, 22.4\% of LLM SATD in our study required entirely new categories. Their Configuration Debt captured hyperparameter tuning but not external API dependencies (Model-Stack Workaround Debt), and their taxonomy models static ML pipelines but not dynamic ecosystem changes (Model Dependency Debt). This move from minor extensions to many new categories reflects a fundamental architectural gap between LLM and traditional ML development.

As a result, many SATD instances labeled as design issues are not OO violations but architectural problems, such as misplaced responsibilities (``\texttt{TODO(matt): move into LLMConfig}'') or inconsistent data handling (``\texttt{TODO: Clean this up to only use one type}''). To account for this, we broaden Design Debt to cover any architectural, procedural, or structural deficiencies that harm maintainability. This expanded definition captured 14.6\% of classified SATD, making it the most common category in our sample.
We also found that traditional Workaround Debt assumes a unified software stack under developer control. LLM systems violate this assumption because they depend on external model stacks that often lack features or contain bugs~\cite{llm_challenges_2025}. Comments such as ``\texttt{HACK: No official tokenizer available for Claude 3}'' and ``\texttt{This is a workaround for the TPU SPMD mode}'' illustrate compensations for external infrastructure deficiencies rather than internal code issues.

\subsubsection{New Debt Categories in the LLM Ecosystem}

Beyond taxonomy extensions, we identify three categories of SATD specific to LLM development, representing 22.4\% of classified instances.
\textbf{Model-Stack Workaround Debt} (6.3\%) represents hacks and compatibility fixes for external ML infrastructure limitations. This category, distinct from traditional workarounds, reflects the reality that LLM developers must constantly adapt to bugs and missing features in fast-paced model APIs, tokenizers, and runtime environments they cannot control.
\textbf{Model Dependency Debt} (11.2\%) captures deferred updates waiting for upstream library evolution. Comments such as ``\texttt{TODO: Remove once accelerate is updated}'' and ``\texttt{TODO: will remove after torch xpu 2.9 supports uuid}'' reveal a dependency-driven debt pattern where code remains provisional until external frameworks provide required features. Unlike traditional dependency management, resolution depends entirely on external release cycles.
\textbf{Performance Optimization Debt} (4.9\%) represents deferred hardware-specific optimizations requiring specialized expertise. Comments reference GPU kernel engineering (``\texttt{TODO: optimize with triton kernels}''), memory layout adjustments, and \textit{FlashAttention} integration (an attention mechanism optimization for transformers). This differs from traditional performance debt, which involves algorithmic improvements, as it requires deep knowledge of hardware acceleration and kernel programming that is often absent from application teams.

The prevalence of LLM-specific debt categories shows how ecosystem volatility shapes technical debt in modern AI systems. The dominance of Model-Stack Workaround Debt, the largest LLM-specific category, indicates that external infrastructure instability drives much of this debt. Developers expend significant effort compensating for limitations in APIs and frameworks beyond their control, a pattern less common in traditional software where most dependencies are stable.
Despite these novel challenges, traditional categories remain prevalent. Requirements Debt (11.9\%) and Code Debt (10.6\%) persist alongside LLM-specific issues, showing that core software engineering challenges transcend paradigm shifts. The high proportion of ``\texttt{Undefined}'' SATD (comments containing only ``\texttt{TODO}'') suggests that debt documentation practices lag behind system complexity.

Table~\ref{tab:satd-taxonomy} presents our complete extended taxonomy. This framework provides researchers with categories necessary for analyzing SATD in contemporary AI systems while offering practitioners a vocabulary for discussing and prioritizing different types of technical debt. The coexistence of traditional and novel debt types suggests that managing LLM technical debt requires both established software engineering practices and new strategies tailored to ecosystem dependencies and performance optimization challenges.

\begin{table*}[t]
\centering
\caption{Extended SATD taxonomy: Original categories (Bavota \& Russo)~\cite{bavota_russo}, prior extensions (Bhatia \textit{et al.}~\cite{bhatia2025empirical}), and new LLM-specific categories introduced in this study (* refers to debt types introduced by this study)}
\vspace{-3pt}
\label{tab:satd-taxonomy}
\renewcommand{\arraystretch}{1.1}
\resizebox{\textwidth}{!}{
\begin{tabular}{|l|p{2.9cm}|p{5.75cm}|p{5cm}|}
\hline
\textbf{Type} & \textbf{Subtype} & \textbf{Description} & \textbf{Example} \\
\hline
\hline
\multicolumn{4}{|c|}{\cellcolor{gray!20}\textbf{Requirements Debt} \textit{(Bavota \& Russo)}} \\
\hline
Functional & Improvement & Comments indicating improvements to existing functionality & ``Network library could be added too.'' \\
\cline{2-4}
Functional & New Feature & Missing features deferred for future implementation & ``TODO handle attention histories.'' \\
\cline{2-4}
Non-Functional & Performance & Improving speed, efficiency, or responsiveness & ``TODO slow loading of encoder may be due to device.'' \\
\hline
\hline
\multicolumn{4}{|c|}{\cellcolor{gray!20}\textbf{Configuration Debt} \textit{(Bhatia et al.)}} \\
\hline
Configuration & -- & Uncertain, temporary, or suboptimal configuration choices & ``300 iterations seems good enough but you can train longer.'' \\
\hline
\hline
\multicolumn{4}{|c|}{\cellcolor{gray!20}\textbf{Code Debt} \textit{(Bavota \& Russo)}} \\
\hline
Code Quality & Low Internal Quality & Poor structure, readability, or maintainability & ``TODO change format of formatted\_preds.'' \\
\cline{2-4}
Code Quality & Refactoring & Calls for structural cleanup, simplification, or removal of dead code & ``TODO make this code simpler.'' \\
\cline{2-4}
Code Quality & Workaround (Traditional) & Temporary or ad-hoc hack used to bypass missing logic & ``Naive approach—probably not robust.'' \\
\hline
\hline
\multicolumn{4}{|c|}{\cellcolor{gray!20}\textbf{LLM-Specific Debt} \textit{(This Study)}} \\
\hline
Model-Stack Workaround* & -- & Workarounds for missing/unstable features in tokenizers, model runtimes, or backends & ``HACK: No official tokenizer available for Claude 3.'' \\
\cline{2-4}
Model Dependency* & -- & Deferred upgrades to external ML libraries or frameworks & ``TODO: Remove once accelerate is updated.'' \\
\cline{2-4}
Performance Optimization* & -- & Deferring GPU/kernel/Triton/attention mechanism optimizations & ``TODO: optimize with Triton kernels.'' \\
\hline
\hline
\multicolumn{4}{|c|}{\cellcolor{gray!20}\textbf{Design Debt} \textit{(Bavota \& Russo)}} \\
\hline
Design & Design Patterns (OO) & Violations of classical design principles & ``Should this method be made more general?'' \\
\cline{2-4}
Design & Expanded Design\textsuperscript{*} & Architectural or pipeline-level design issues (e.g., misplaced components) & ``TODO: move this into LLMConfig.'' \\
\hline
\hline
\multicolumn{4}{|c|}{\cellcolor{gray!20}\textbf{Defect Debt} \textit{(Bavota \& Russo)}} \\
\hline
Defect & Known Defect & Acknowledged incorrect behavior requiring a fix & ``TODO unintended side effect on input.'' \\
\hline
\hline
\multicolumn{4}{|c|}{\cellcolor{gray!20}\textbf{Testing Debt} \textit{(Bhatia et al.)}} \\
\hline
Testing & Inadequate Testing & Missing or incomplete tests & ``TODO add test for conv2d.'' \\
\hline
\hline
\multicolumn{4}{|c|}{\cellcolor{gray!20}\textbf{Other Debt Categories} \textit{(Bavota \& Russo)}} \\
\hline
Undefined & -- & SATD comment with no descriptive content & ``TODO'' \\
\cline{2-4}
On Hold & -- & Debt comment remains though work is complete & ``TODO maxlen'' (already implemented) \\
\cline{2-4}
Documentation & -- & Missing or incomplete documentation & ``TODO add logger info.'' \\
\hline
\end{tabular}
}
\vspace{-11pt}
\end{table*}

\vspace{3pt}
\noindent\textbf{Practical Implication.}
The distribution of SATD types carries clear engineering implications. The high share of external-dependency debt (Model-Stack Workaround $+$ Model Dependency: 11.1\%) signals a need for dedicated resources to track upstream changes and maintain compatibility layers. The presence of Performance Optimization Debt suggests teams must either invest in specialized performance engineering or accept that some optimizations will remain deferred. Most importantly, the broadened notion of Design Debt in non-OO systems calls for rethinking how architectural quality is evaluated and preserved in modern AI codebases.

\begin{tcolorbox}[colback=gray!10, colframe=gray!50, boxrule=0.5pt, left=2pt, right=2pt, top=2pt, bottom=2pt]
\textbf{Answer to RQ2:} LLM repositories exhibit 22.4\% novel SATD types: Model-Stack Workaround (6.3\%), Model Dependency (11.2\%), and Performance Optimization Debt (4.9\%). While traditional debt categories like Design (11.2\%) and Requirements (23.1\%) remain dominant, these LLM-specific types reflect unique challenges in external infrastructure dependencies and hardware optimization not present in traditional or pre-LLM ML systems.
\end{tcolorbox}

\subsection{RQ3: Distribution of SATD Across LLM Development Pipeline Stages}

\begin{figure}
    \centering
    \includegraphics[width=0.97\linewidth]{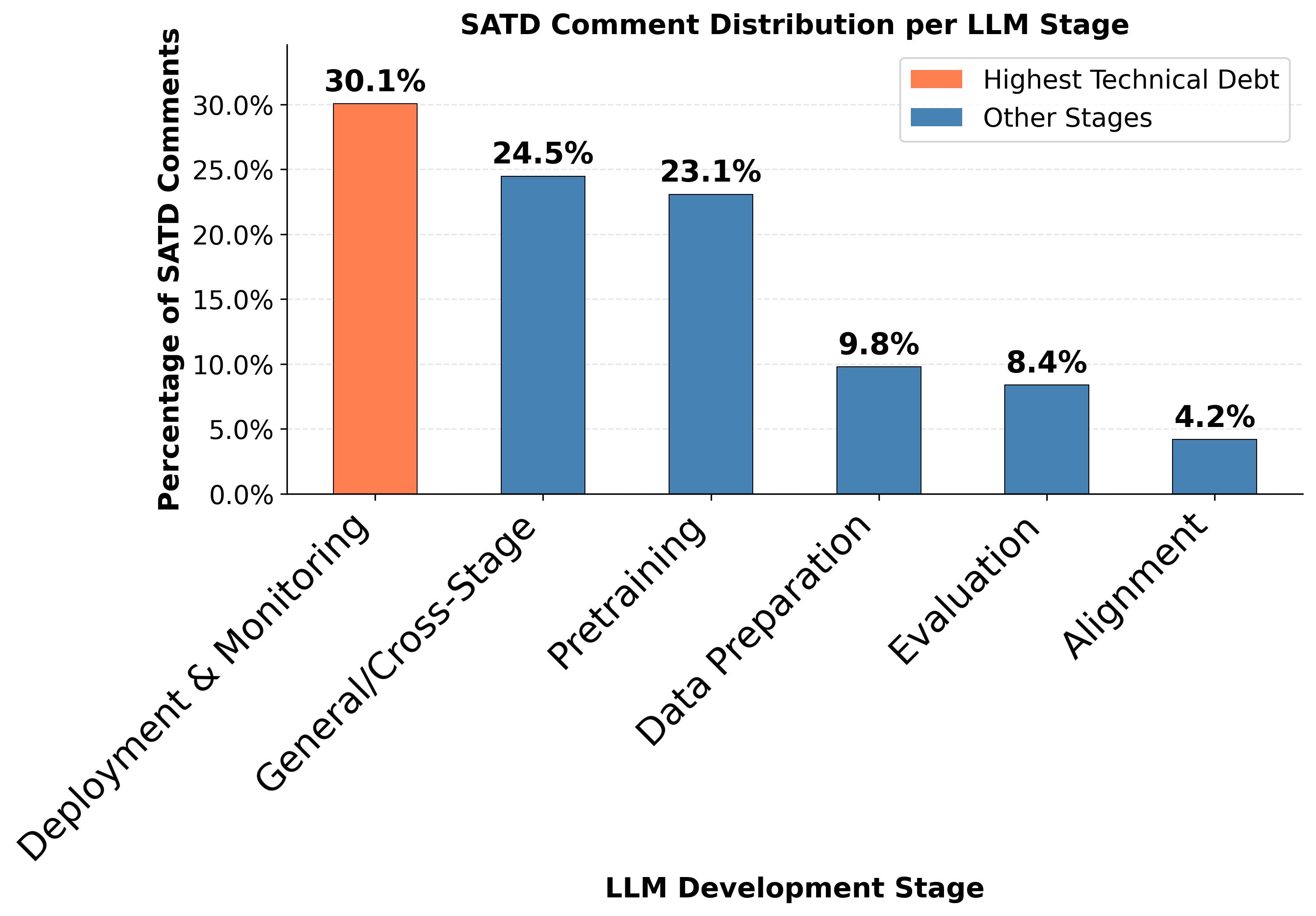}
    \vspace{-3pt}
    \caption{\textit{Distribution of SATD Comments Across LLM Development Pipeline Stages}}
    \label{fig:stage_dis}
    \vspace{-13pt}
\end{figure}

We map SATD instances to the LLM development pipeline stages defined by Hu \textit{et al.}~\cite{hu2024llmcharacterization} to understand where technical debt concentrates in modern LLM workflows. Figure~\ref{fig:stage_dis} presents the distribution across five pipeline stages plus a General category for shared infrastructure not tied to any specific pipeline stage.
The concentration of SATD in \textit{Deployment and Monitoring} (30\% of stage-specific debt) highlights the operational difficulty of serving LLMs at scale. This stage requires sophisticated optimizations, such as quantization, kernel tuning, memory scheduling, and parallel serving, thus creating many opportunities for provisional solutions. Some comments acknowledge deferred optimizations (``\texttt{TODO: implement dynamic batching}''), temporary workarounds for serving constraints (``\texttt{HACK: fixed batch size until memory profiling complete}''), and configuration shortcuts taken under production pressure.

This distribution shows that operational infrastructure, rather than model logic, drives debt accumulation; instead, operational infrastructure emerges as the primary debt source. The complexity of operationalizing LLMs, balancing latency, throughput, and resource utilization, forces teams to accept provisional solutions that often become permanent, with many SATDs referencing integration challenges with serving frameworks and deferred optimizations awaiting specialized expertise.

Pretraining (23\%) is the second-largest concentration of debt, with SATD referencing missing checkpointing logic, incomplete parallelization, and deferred memory optimizations. The resource-intensive nature of pretraining creates pressure to maintain forward progress rather than address technical debt. 

Alignment (4\%) and Evaluation (8\%) show minimal SATD, consistent with their short-running jobs and modest resource requirements. Debt in these stages involves missing features (``\texttt{TODO: add ROUGE metric}'') rather than architectural issues. This pattern suggests that modularity and operational simplicity act as natural barriers to debt accumulation, as evaluation and alignment scripts can be modified independently, making debt both less likely to accumulate and easier to address.

Data Preparation (10\%) occupies a middle position, with complex ETL processes and quality filters accumulating moderate debt through incomplete preprocessing and temporary edge-case solutions. The General category (24\%) captures cross-cutting infrastructure debt (logging, configuration, monitoring) that pipeline-focused analyses often overlook. Comments like ``\texttt{TODO: unify configuration format across modules}'' indicate that shared components are major debt sources impacting overall maintainability.

\vspace{3pt}
\noindent\textbf{Practical Implication.}
Debt concentrated in deployment and pretraining should guide most refactoring efforts, particularly in deployment infrastructure, where debt directly affects production reliability. Low debt in alignment and evaluation permits faster iteration with minimal technical risk. The prevalence of general infrastructure debt shows that shared components require ongoing maintenance rather than being treated as secondary. This distribution also informs risk assessment: deployment debt poses immediate operational risks, pretraining debt undermines reproducibility and training stability, and evaluation or alignment debt is generally less critical. Teams should prioritize resolving debt where shortcuts trigger systemic failures rather than isolated feature limitations.

\begin{tcolorbox}[colback=gray!10, colframe=gray!50, boxrule=0.5pt, left=2pt, right=2pt, top=2pt, bottom=2pt]
\textbf{Answer to RQ3:} SATD is concentrated in the Deployment/Monitoring (30\%) and Pretraining (23\%) stages of the LLM pipeline, with minimal accumulation in Alignment (4\%) and Evaluation (8\%). This suggests that operational complexity and infrastructure, rather than model logic or algorithms, drive SATD accumulation in LLM systems.
\end{tcolorbox}
\subsection{RQ4: Temporal SATD Characteristics Across Repository Types}
\vspace{-2pt}

We analyze the complete git history of all repositories to understand when SATD appears and how long it persists across different software paradigms. Our survival analysis methodology (detailed below) tracks both introduction timing and persistence patterns. Table~\ref{tab:rq4_metrics} summarizes the temporal characteristics from our analysis of over 5 million comment events.\vspace{-2pt}

\begin{tcolorbox}[title=Technical Details: Survival Analysis Methodology, fonttitle=\bfseries, left=2pt, right=2pt, top=2pt, bottom=2pt]
\small
\textbf{Project-Level Introduction:} Time from first commit to first SATD appearance:
$T_{\text{project}} = T_{\text{first-SATD}} - T_{\text{project-start}}$

\vspace{3pt}
\textbf{File-Level Introduction:} Time from file creation to first SATD in that file:
$T_{\text{file}} = T_{\text{SATD-intro}} - T_{\text{file-create}}$

\vspace{3pt}
\textbf{SATD Survival:} Duration from introduction to removal (or censoring):
$T_{\text{survival}} = T_{\text{removal}} - T_{\text{introduction}}$

\vspace{3pt}
\textbf{Statistical Analysis:} Kaplan-Meier estimation with log-rank tests for group comparisons. Censoring applied to unresolved SATD at study endpoint.
\end{tcolorbox}
\vspace{-2pt}

\begin{table}[ht]
\centering
\caption{SATD temporal characteristics (* refer to total additions and removals across all commits)}
\vspace{-3pt}
\label{tab:rq4_metrics}
\begin{tabular}{p{3.9cm}rrr}
\hline
\textbf{Metric} & \textbf{LLM} & \textbf{ML} & \textbf{Non-ML} \\
\hline
    Projects Analyzed         & 159       & 159       & 159      \\
    Comment Events*           & 3,713,429 & 1,212,899 & 661,120  \\
    Total SATD Detected       & 91,046    & 29,384    & 14,805   \\
    \hline
    Median Introduction (days) & 492       & 204       & 1,005    \\
    Median Survival (days)    & 553       & 401       & 776      \\
    Removal Rate (\%)         & 49.1      & 53.9      & 55.8     \\
\hline
\end{tabular}
\vspace{-3pt}
\end{table}

\subsubsection{LLM Projects: Delayed Introduction, Persistent Accumulation}

LLM repositories exhibit a two-phase development pattern. They remain SATD-free for a median of 492 days, far longer than ML projects (204 days) but shorter than traditional software (1,005 days), reflecting a prolonged infrastructure-building phase before complexity forces shortcuts. Once SATD appears, it persists: with a 553-day median lifespan and the lowest removal rate (49.1\%), debt becomes structurally embedded. Tight coupling across prompts, APIs, and orchestration layers creates architectural lock-in, making refactoring costly, unlike ML systems, where experimental components can be replaced with far less risk. File-level analysis (Table~\ref{tab:rq4_file_intro}) confirms this: LLM files wait 1,144 days (median) before first SATD appears, suggesting careful early development followed by inevitable compromise as complexity grows.

\begin{table}[ht]
\centering
\vspace{-2pt}
\caption{File-level SATD introduction timing}
\vspace{-3pt}
\label{tab:rq4_file_intro}
\begin{tabular}{p{2cm}rrrr}
\hline
\textbf{Cohort} & \textbf{Total} & \textbf{Files w/} & \textbf{\% Files} & \textbf{Median Days} \\
                & \textbf{Files} & \textbf{SATD}      & \textbf{w/ SATD}  & \textbf{to First SATD} \\
\hline
LLM             & 84,019         & 38,417            & 45.7\%            & 1,144 \\
ML              & 20,726         & 11,060            & 53.4\%            & 441 \\
Non-ML          & 20,398         & 10,359            & 50.8\%            & 2,152 \\
\hline
\end{tabular}
\vspace{-1pt}
\end{table}

\subsubsection{Repository Maturation: Understanding the $44x$ Shift}

Our results show a major temporal shift in ML repositories: the median time to first SATD increased from 10 days (Bhatia \textit{et al.}, 2021) to 441 days in our 2025 analysis, a $44x$ increase. This reflects repository maturation rather than methodological differences. Around 2021, ML repositories were in experimental phases with rapid file creation and immediate SATD as developers prototyped~\cite{obrien2022shades}. By 2025, many experimental files had been removed or stabilized, leaving a mature core where SATD appears only after multiple development cycles. This demonstrates that SATD behavior evolves over the project lifecycle. The result highlights a methodological challenge: cross-sectional studies capture only a snapshot and can misrepresent long-term debt patterns. SATD analysis therefore requires a temporal perspective, as the same repository can behave very differently at different stages of maturity.

\subsubsection{The Permanence of Foundational Debt}

While overall removal rates appear reasonable (49-56\%), file-level analysis reveals a startling pattern: the first SATD introduced into a file is almost never removed (Table~\ref{tab:rq4_file_removal}). Removal rates below 5\% across all repository types indicate that initial technical debt becomes permanent, embedded in architectural decisions too costly to revisit. This contrast between comment-level removal ($\approx$50\%) and first-file-SATD removal ($<$5\%) reveals two modes of debt management. Developers actively clean recent, localized debt through routine maintenance, but accommodate rather than eliminate foundational compromises. The temporal placement of SATD within a file matters as much as its content. Early debt becomes structural, while later additions remain more tractable.

\begin{table}[ht]
\centering
\caption{First SATD removal rates by file}
\vspace{-3pt}
\label{tab:rq4_file_removal}
\begin{tabular}{lrrr}
\hline
\textbf{Cohort} & \textbf{Files w/ SATD} & \textbf{First SATD Removed} & \textbf{\% Removed} \\
\hline
LLM     & 38,417 & 1,628 & 4.2\% \\
ML      & 11,060 & 311   & 2.8\% \\
Non-ML  & 10,359 & 432   & 4.2\% \\
\hline
\end{tabular}
\vspace{-10pt}
\end{table}

\subsubsection{Survival Curves: Visualizing Paradigm Differences}

\begin{figure}
    \centering
    \includegraphics[width=1\linewidth]{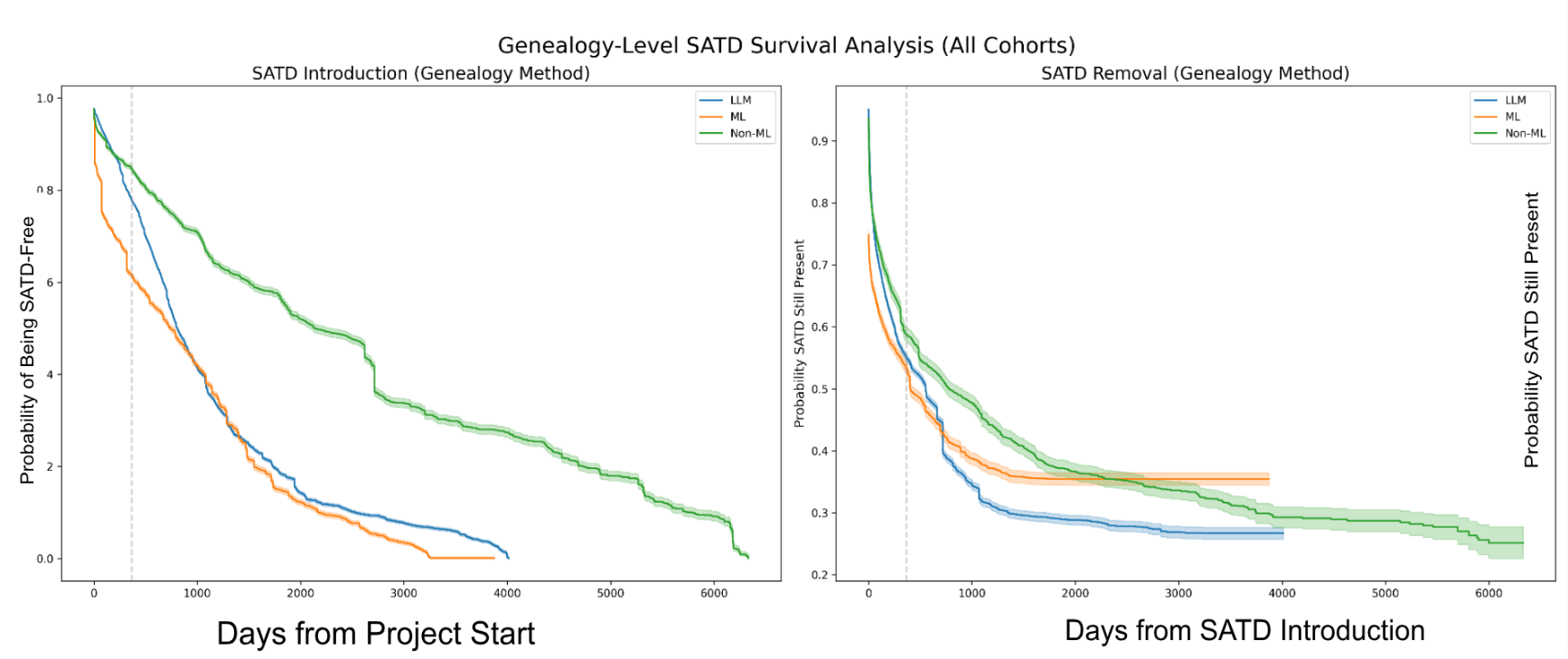}
    \caption{\textit{Kaplan–Meier survival curves for SATD introduction (left) and removal (right).}}
    \label{fig:rq4_survival}
    \vspace{-8pt}
\end{figure}

Figure~\ref{fig:rq4_survival} presents Kaplan-Meier~\cite{kaplan_meier_1958} survival curves demonstrating distinct patterns across repository types. ML projects show steepest SATD introduction (left panel) but also fastest removal (right panel), reflecting rapid experimentation with active cleanup. LLM projects exhibit delayed introduction but flatter removal curves, indicating debt that accumulates slowly but persists. Non-ML projects show gradual introduction and removal, consistent with stable development practices. Curves quantify how development paradigms shape debt dynamics. ML's experimental nature drives early debt that teams actively manage. Although the complexity of LLM infrastructure can delay accumulating debt, it can restrict architectural flexibility. Traditional software's stability produces gradual, manageable debt accumulation. These patterns suggest that debt management strategies must adapt to paradigm-specific temporal characteristics rather than applying universal approaches.

\vspace{3pt}
\noindent\textbf{Practical Implication.}
Foundational debt durability underscores early design choices: shortcuts during initial file creation rarely get revisited, necessitating stricter review for new files. For LLM projects, the delayed-but-persistent pattern suggests front-loading architectural investment to extend the debt-free phase, as later debt becomes permanent. Given $<$5\% removal rates, prevention outweighs cleanup planning. Repository age affects interpretation: high SATD in young projects reflects experimentation, while spikes in mature projects signal architectural degradation. Age-adjusted metrics provide better technical health indicators than raw SATD counts.

\begin{tcolorbox}[colback=gray!10, colframe=gray!50, boxrule=0.5pt, left=2pt, right=2pt, top=2pt, bottom=2pt]
\textbf{Answer to RQ4:} LLM repositories stay debt-free longer than ML projects (492 vs. 204 days), but once technical debt appears, it becomes hard to remove. Nearly half of SATD remains unresolved, with over 95\% of the first SATD introduced in a file never removed. This suggests LLM projects start clean but quickly reach architectural lock-in, though low removal rates may reflect the younger age of LLM repositories compared to mature ML projects.
\end{tcolorbox}

\subsection{RQ5: Predictors of Long-Lasting SATD}

Following Bhatia \textit{et al.}~\cite{bhatia2025empirical}, we train Random Forest classifiers to identify what likely makes some technical debts last longer than others. We analyze commit-level features (lines added, lines modified, files changed) and file-level features (lines of code, complexity metrics, token counts) to predict debt persistence. To find predictive patterns, the models classify each debt as (1) long-lasting debt (surviving longest 25\%), or (2) quickly-removed debt (removed fastest 25\%). Table~\ref{tab:rq5_performance} presents the model performance and top predictors for each repository type.

\begin{table}[t]
\centering
\caption{Model performance and most important predictors of long-lasting SATD}
\vspace{-3pt}
\label{tab:rq5_performance}
\begin{tabular}{llcc}
\hline
\textbf{Cohort} & \textbf{Top Predictors} & \textbf{Importance} & \textbf{Accuracy} \\
\hline
\multirow{3}{*}{\textbf{LLM}} 
    & Lines modified in commit & 0.125 & \multirow{3}{*}{0.82} \\
    & Code tokens in file & 0.115 & \\
    & Lines of code in file & 0.111 & \\
\hline
\multirow{3}{*}{\textbf{ML}} 
    & Lines added in commit & 0.190 & \multirow{3}{*}{0.96} \\
    & Historical file changes & 0.150 & \\
    & Lines modified in commit & 0.150 & \\
\hline
\multirow{3}{*}{\textbf{Non-ML}} 
    & Code tokens in file & 0.118 & \multirow{3}{*}{0.74} \\
    & Cyclomatic complexity & 0.118 & \\
    & Lines of code in file & 0.116 & \\
\hline
\end{tabular}
\vspace{-9pt}
\end{table}

Our models predict long-lasting debt with 96\% accuracy in ML repositories, 82\% in LLM repositories, and 74\% in traditional software. ML projects follow similar experimental workflows, making debt patterns consistent and predictable. LLM projects mix experimentation with production engineering, creating more varied debt patterns. Traditional software has the most diverse debt sources—user interfaces, databases, business logic—making prediction hardest. By examining feature importance (Table~\ref{tab:rq5_performance}), we observe that the predictors reveal why debt persists across different systems. In ML repositories, large commits (hundreds of lines added or modified) create the most persistent debt. When researchers run large experiments or refactor entire pipelines, the resulting debt often becomes permanent. LLM repositories show a different pattern: debt in large files (by lines of code or token count) persists, reflecting that large inference pipelines and serving infrastructure are difficult to refactor once deployed. Traditional software relies on complexity metrics. When code becomes complex (high cyclomatic complexity) or files grow large, debt becomes harder to understand and fix, thus persists.

A universal pattern emerges: large commits create lasting debt regardless of project type. When teams rush to ship features or fix urgent problems, they create debt that rarely gets cleaned up. Teams can act on this immediately: because commit size strongly predicts persistent debt, tools should flag large commits ($>500$ lines) for extra review. In LLM projects, review should focus on large infrastructure files where debt most likely becomes permanent. Varying prediction accuracy guides tool development. ML teams can use simple commit-size rules with high confidence, while LLM and traditional software teams need sophisticated multi-factor approaches. SATD tools must adapt to each domain rather than rely on universal approaches.\vspace{-2pt}

\begin{tcolorbox}[colback=gray!10, colframe=gray!50, boxrule=0.5pt, left=2pt, right=2pt, top=2pt, bottom=2pt]
\textbf{Answer to RQ5:} Large commits consistently predict long-lasting SATD across all repository types, but prediction accuracy varies by domain specialization (LLM: 82\%, ML: 96\%, Non-ML: 74\%).
\end{tcolorbox}

\section{Implications}
\label{sec:Implications}
\vspace{-1.5pt}
\noindent\textbf{For Practitioners: prevent debt early, especially in critical stages.}
The first SATD introduced in a file has a $<$5\% removal rate, making early architectural decisions effectively permanent; developers should therefore enforce stricter reviews for new files. Large commits ($>$500 lines or $>$10 files) correlate with $3–4$ times longer debt persistence and should trigger CI/CD warnings. Debt concentrates in deployment, monitoring, and pretraining (RQ3), warranting targeted refactoring. Since 22.4\% of LLM debt comes from external issues like model-stack workarounds (RQ2), teams should track upstream changes and maintain compatibility layers.

\vspace{1.2pt}
\noindent\textbf{For Tool Builders: paradigm-specific detection and prevention.}  
RQ2 identifies three LLM-specific debt categories that existing SATD detectors, trained on traditional ML or non-ML comments, will miss. Therefore, detectors should be retrained on LLM-specific taxonomies. RQ5 shows SATD prediction accuracy differs by paradigm (LLM: 0.82, ML: 0.96, non-ML: 0.74), motivating paradigm-specific tooling. RQ3 suggests stage-aware warnings should prioritize debt in deployment and pretraining code over evaluation or alignment~scripts.

\vspace{1.2pt}
\noindent\textbf{For Researchers: multi-level, temporal analysis is essential.}  
RQ4 challenges single-snapshot SATD studies through the $44x$ increase in file-level SATD introduction time (10 days in 2021 vs. 441 days in 2025), highlighting how technical debts evolve with repository age. Researchers should treat repository maturity as a confounding variable and use longitudinal studies to capture insights unobservable in single snapshots. The contrasting removal rates (54\% at comment-level versus 3-5\% for first-file SATD) reveal that analysis granularity affects conclusions about debt management effectiveness. Future SATD research should adopt multi-level frameworks reporting metrics at comment, file, and project levels while distinguishing initial from subsequent debt.

\vspace{1.3pt}
\noindent\textbf{For the broader SE community: LLM development as a hybrid paradigm.}
Across all five research questions, LLM development emerges as a distinct hybrid paradigm. RQ1 shows comparable debt rates to ML (3.95\% vs. 4.10\%), but RQ4 reveals different timing (2.4x longer debt-free periods followed by rapid accumulation). RQ2 attributes debt to external ecosystem volatility; RQ3 shows it concentrates in operational infrastructure rather than model logic. Thus, LLM development differs from both traditional software engineering (which assumes stable dependencies) and classical ML (which assumes experimental workflows). Converging debt levels despite differing temporal patterns suggests technical debt is fundamental to software evolution, not AI-specific.

\section{Threats to Validity}
\label{sec:Threats}
\vspace{-1pt}
\noindent\textbf{Construct Validity.}
Our analysis treats SATD as a proxy for technical debt, excluding non-admitted, implicit, and systemic debt, and thus likely underestimates total debt. The NLP-based detector, previously reporting 0.82 F1-score, yielded 44\% false positives in our dataset, requiring manual filtering that may introduce selection bias. The taxonomy extension relies on 377 manually classified instances with 85.7\% initial agreement; the remaining 14.3\% suggest ambiguous, LLM-specific debt boundaries. Expanding Design Debt beyond OO architectures may hinder comparison with prior OO-focused studies.

\vspace{1.5pt}
\noindent\textbf{Internal Validity.}
Repository age confounds temporal comparisons: LLM repositories (post-2022) are younger than ML repositories (2015–2021), making it unclear whether differences reflect age or paradigm shifts. The $44x$ increase in ML file-level introduction time (10 days in 2021 vs. 441 days in 2025) shows how age alters debt patterns, yet young LLM projects are compared with mature ML projects. Survival analysis censors unresolved SATD at study endpoints, likely inflating persistence metrics for recent commits. Random Forest models may overlook LLM-specific factors such as prompt quality, API stability, or RAG complexity. Focusing only on the master branch omits debt in experimental branches.

\vspace{1.5pt}
\noindent\textbf{External Validity.}
Findings are limited to open-source Python projects on GitHub; generalizability to other languages, closed-source systems, or different contexts is uncertain. LLM repository selection focused on post-ChatGPT projects using specific frameworks (LangChain, LlamaIndex) and may not represent all LLM development patterns, particularly enterprise systems or research codebases developing novel architectures. The 159-repository sample per category, while statistically powered for prevalence comparisons, may miss rare debt patterns in specialized domains (e.g., medical or financial LLMs). Pipeline stage mapping relies on manual classification of file paths and content, potentially misclassifying cross-cutting concerns or hybrid components. GitHub projects may overrepresent Western, English-speaking, startup contexts compared to industrial AI development elsewhere.

\section{Conclusion}
\label{sec:Conclusion}
\vspace{-1pt}
This study presents the first systematic replication and extension of SATD analysis in the LLM era. Applying Bhatia \textit{et al.}'s methodology to 477 LLM, ML, and non-ML repositories reveals both continuity and evolution in technical debt patterns. Our contributions include: (1) evidence that SATD prevalence gaps across software categories are narrowing, (2) identification of three LLM-specific debt types extending existing taxonomies, (3) analysis of repository maturation effects, and (4) a reproducible framework for longitudinal SATD analysis. SATD pattern convergence across paradigms suggests technical debt is inherent to software evolution, not a sign of poor practice. Understanding LLM-specific debt accumulation is essential for sustainable AI development.
Future work should develop CI/CD bots to flag high-risk pull requests and prevent lasting debt. Researchers should extend the SATD taxonomy using larger datasets or domain-specific LLMs to uncover additional debt patterns. Qualitative studies with LLM practitioners could clarify how they weigh debt tradeoffs and the challenges of resolving foundational debt.

\section*{Acknowledgment}
\vspace{-1pt}
This work is funded by the Natural Sciences and Engineering Research Council of Canada (NSERC): RGPIN-2023-05440 and RGPIN-2025-05897.

\balance
\bibliographystyle{IEEEtran}
\bibliography{references}

\end{document}